\documentclass[pra,amsmath,amssymb,two column]{revtex4}
\usepackage{mathrsfs}
\usepackage{amssymb}

\input epsf.tex
\usepackage{graphicx}
\usepackage{amsthm}
\usepackage{enumitem}

\begin{document}

\title{Decoy-state measurement-device-independent quantum key distribution based on the Clauser-Horne-Shimony-Holt inequality}
\author{Chun-Mei Zhang$^{1,2}$, Mo Li$^{1}$, Hong-Wei Li$^{1,2,a}$, Zhen-Qiang Yin$^{1,b}$,  Dong Wang$^{1}$,  Jing-Zheng Huang$^{1}$, Yun-Guang Han$^{1}$, Man-Li Xu$^{1}$, Wei Chen$^{1}$, Shuang Wang$^{1}$, Patcharapong Treeviriyanupab$^{3}$, Guang-Can Guo$^{1}$, Zheng-Fu Han$^{1}$ }

 \affiliation
 {$^1$ Key Laboratory of Quantum Information, CAS, and Synergetic Innovation Center of Quantum Information \& Quantum Physics, University of Science and Technology of China, Hefei 230026, China\\
$^2$State Key Laboratory of Networking and Switching Technology, Beijing University of Posts and Telecommunications, Beijing 100876, China\\
$^3$Department of Computer Science, Faculty of Science and Technology, Phranakhon Rajabhat University, Bangkok 10220, Thailand}

\begin{abstract}
The measurement-device-independent quantum key distribution (MDI-QKD) protocol is proposed to remove the detector side channel attacks, while its security relies on the assumption that the encoding systems are perfectly characterized. In contrast, the MDI-QKD protocol based on the Clauser-Horne-Shimony-Holt inequality (CHSH-MDI-QKD) weakens this assumption, which only requires the quantum state to be prepared in the two-dimensional Hilbert space and the devices are independent. In experimental realizations, the weak coherent state, which is always used in QKD systems due to the lack of an ideal single photon source, may be prepared in the high-dimensional space. In this paper, we investigate the decoy-state CHSH-MDI-QKD protocol with $s(3 \le s \le 5)$ intensities, including one signal state and $s-1$ decoy states, and we also consider the finite-size effect on the decoy-state CHSH-MDI-QKD protocol with five intensities. Simulation results show that this scheme is very practical.
\end{abstract}
\maketitle

{\it Introduction  - } In principle, quantum key distribution (QKD) \cite{BB84} allows two distant legitimate parties Alice and Bob to share unconditional secret keys, even in the presence of an eavesdropper Eve. Nevertheless, practical implementations of QKD systems are usually composed of imperfect devices, which makes it vulnerable to be attacked by Eve \cite{attack1,attack2,attack3,attack4,attack5}. To avoid all possible loopholes existing in real-life QKD systems, the device-independent QKD (DI-QKD) protocol \cite{DI} is proposed, whose security relies on the violation of the Clauser-Horne-Shimony-Holt (CHSH) inequality \cite{CHSH}. However, DI-QKD is impractical with current technology due to the requirement of near-unity detection efficiency and low channel loss. To avoid the detection loophole problem caused by the channel losses, DI-QKD with a local Bell test has been proposed \cite{Lim}, but it cannot overcome the detection loophole problem caused by the limited detection efficiency. Instead of DI-QKD, the measurement-device-independent QKD (MDI-QKD) protocol \cite{Braunstein,Lo} was proposed to remove all detector side channel attacks. Recently, some experimental demonstrations of MDI-QKD have been performed \cite{MDIEx1,MDIEx2,MDI4,MDI5} and thus MDI-QKD has proved to be practical.

However, the security of MDI-QKD relies on the assumption that the encoding systems are fully characterized. To further improve the security of MDI-QKD, several protocols have been proposed \cite{wangxiangbin, MDI2, Hongwei} to relax its assumptions on the encoding systems. Inspired by the spirit of DI-QKD, the MDI-QKD protocol based on the CHSH inequality (CHSH-MDI-QKD) \cite{Hongwei} has been proposed to weaken the assumption of state preparation in MDI-QKD. Interestingly, CHSH-MDI-QKD can overcome the detection loophole problem existing in DI-QKD. It should be noted that CHSH-MDI-QKD requires the state to be prepared in the two-dimensional Hilbert space, and Alice's (Bob's) devices are independent of Eve's. In the CHSH-MDI-QKD protocol, its state preparation can be assumed to be a black box only with a dimension restriction.

The weak coherent state source, which has a Poisson distribution of photon numbers, is used to implement the CHSH-MDI-QKD protocol due to the lack of an ideal single photon source. The nonzero probability of multi-photon pulses in the weak coherent state, which is equivalent to the high-dimensional state preparation, can bring the photon-number-splitting (PNS) attack \cite{PNS1,PNS2} to CHSH-MDI-QKD. Hence, the decoy-state method \cite{decoy1,decoy2} is adopted to combat the multi-photon events in the weak coherent state. In this Brief Report, we investigate the decoy-state CHSH-MDI-QKD protocol with $s(3 \le s \le 5)$ intensities (including one signal state and $s-1$ decoy states) to estimate the yield and the CHSH value of single-photon contributions, and we also consider the finite-size effect on the decoy-state CHSH-MDI-QKD protocol with five intensities. The simulation results show that our scheme is very promising and can be applied to real-life QKD systems.

{\it Decoy-state CHSH-MDI-QKD protocol  - }Let Alice and Bob use weak coherent states as their sources. Let the predetermined intensity set of Alice (Bob)  be $\mu $ ($\nu $), where $\mu $ ($\nu $) has $m_a$ ($m_b$) different intensities denoted as ${\mu _0},{\mu _1}, \cdots ,{\mu _{{m_a} - 1}}$ (${\nu _0},{\nu _1}, \cdots ,{\nu _{{m_b} - 1}}$). Alice prepares her states in the basis set $\{ {A_1},{A_2}\} $, and Bob prepares his states in the basis set $\{ {B_0},{B_1},{B_2}\} $. Note that the security of CHSH-MDI-QKD does not rely on the details of the basis sets, provided that all the bases are in a two-dimensional Hilbert space. For a real-life implementation, we can assume that ${A_1} = Z$, ${A_2} = X$, ${B_0} = Z$, ${B_1} = \frac{{ - Z - X}}{{\sqrt 2 }}$, and ${B_2} = \frac{{ Z - X}}{{\sqrt 2 }}$, where $Z = \left| 0 \right\rangle \left\langle 0 \right| - \left| 1 \right\rangle \left\langle 1 \right|$, and $X = \left| 0 \right\rangle \left\langle 1 \right| + \left| 1 \right\rangle \left\langle 0 \right|$. Assume the state preparation and measurement devices are independent, which can be satisfied in practical scenarios. The decoy-state CHSH-MDI-QKD protocol \cite{Hongwei} runs as follows:
\begin{enumerate}[nosep]
\item[$(1)$] Alice randomly chooses a basis from $\{ Z,X\} $, a bit from $\{ 0,1\} $, and an intensity from $\mu $, and sends the corresponding weak coherent state pulse to the third untrusted party, Eve. Bob randomly chooses a basis from $\{ Z,\frac{{ - Z - X}}{{\sqrt 2 }},\frac{{Z - X}}{{\sqrt 2 }}\} $, a bit from $\{ 0,1\} $, and an intensity from $\nu $, and sends the corresponding state to Eve.
\item[$(2)$] Considering Alice's states in $\{ Z,X\} $, Bob's states in $\{\frac{{ - Z - X}}{{\sqrt 2 }},\frac{{Z - X}}{{\sqrt 2 }}\} $, and Eve's measurement of the projection into the Bell state $\left| {{\psi ^ - }} \right\rangle  = \frac{1}{{\sqrt 2 }}(\left| {HV} \right\rangle  - \left| {VH} \right\rangle )$, Alice and Bob denote the results as ${S_1}$. Similarly, considering Alice's states in $Z$, Bob's states in $Z$, and Eve's measurement of the projection into the Bell states $\left| {{\psi ^ + }} \right\rangle  = \frac{1}{{\sqrt 2 }}(\left| {HV} \right\rangle  + \left| {VH} \right\rangle )$ and $\left| {{\psi ^ - }} \right\rangle  = \frac{1}{{\sqrt 2 }}(\left| {HV} \right\rangle  - \left| {VH} \right\rangle )$, Alice and Bob denote the results as ${S_2}$. Note that, in ${S_2}$, Alice or Bob should flip her or his bits so that their bit strings are correctly correlated.
\item[$(3)$] Alice and Bob estimate the CHSH value ${g_{11}}$ of single-photon contributions according to ${S_1}$, and estimate the yield $Y_{11}^{ZZ}$ of single-photon contributions according to ${S_2}$, where ${S_1}$ and ${S_2}$ consist of all combinations of Alice's and Bob's intensities. Then, they perform key reconciliation and privacy amplification to get final secret keys.
\end{enumerate}

Theoretically, after the quantum communication phase, Alice and Bob can get a series of decoy-state equations which can be expressed as
\begin{equation}
Q_{{\mu _k}{\nu _l}}^{ZZ} = \sum\nolimits_{m,n = 0}^\infty  {P_{mn}^{{\mu _k}{\nu _l}}Y_{mn}^{ZZ}} ,
\end{equation}
\begin{equation}
\sum\nolimits_{i,j = 0}^1 {{a_{ij}}Y_{{\mu _k}{\nu _l}}^{ij,w}}  = \sum\nolimits_{m,n = 0}^\infty  {P_{mn}^{{\mu _k}{\nu _l}}\sum\nolimits_{i,j = 0}^1 {Y_{mn}^{ij,w}C_{mn}^w} } ,
\end{equation}
and
\begin{equation}
\sum\nolimits_{i,j = 0}^1 {Y_{{\mu _k}{\nu _l}}^{ij,w} = } \sum\nolimits_{m,n = 0}^\infty  {P_{mn}^{{\mu _k}{\nu _l}}\sum\nolimits_{i,j = 0}^1 {Y_{mn}^{ij,w}} } ,
\end{equation}
where $\mu_k$ ($\nu_l$) is the intensity of Alice's (Bob's) weak coherent state source, $m$ ($n$) denotes that Alice (Bob) sends out a weak coherent pulse with $m$ ($n$) photons, $i$ ($j$) is the eigenstate in Alice's (Bob's) basis, $ZZ$ denotes the combination of Alice's basis $Z$ and Bob's basis $Z$, $w$ is the combination of Alice's and Bob's basis ($w \in \{ Z\frac{{ - Z - X}}{{\sqrt 2 }},X\frac{{ - Z - X}}{{\sqrt 2 }},X\frac{{Z - X}}{{\sqrt 2 }},Z\frac{{Z - X}}{{\sqrt 2 }}\} $, for simplicity, which is also denoted as $w \in \{ QS,RS,RT,QT\} $ in the same sequence \cite{Nielsen}), ${{a_{ij}}}$ is the constant parameter (${a_{00}} = 1,{a_{01}} =  - 1,{a_{10}} =  - 1,{a_{11}} = 1$), ${P_{mn}^{{\mu _k}{\nu _l}}}$ is the probability of Alice sending out $m$ photons with the intensity $\mu_k$ and Bob sending out $n$ photons with the intensity $\nu_l$, ${Y_{{\mu _k}{\nu _l}}^{ij,w}}$ is the yield of the eigenstates $i$ and $j$ with the intensities $\mu_k$ and $\nu_l$ in the basis combination $w$, ${Y_{mn}^{ij,w}}$ is the yield of the eigenstates $i$ and $j$ with the photons $m$ and $n$ in the basis combination $w$, $Q_{{\mu _k}{\nu _l}}^{ZZ} = \frac{1}{4}\sum\nolimits_{i,j = 0}^1 {Y_{{\mu _k}{\nu _l}}^{ij,ZZ}} $, $Y_{mn}^{ZZ} = \frac{1}{4}\sum\nolimits_{i,j = 0}^1 {Y_{mn}^{ij,ZZ}} $, and $C_{mn}^w = (\sum\nolimits_{i,j = 0}^1 {{a_{ij}}Y_{mn}^{ij,w}} )/(\sum\nolimits_{i,j = 0}^1 {Y_{mn}^{ij,w}} )$.

As suggested in \cite{MDI1,MDI3}, without loss of security and accuracy, there is no need to consider an infinite number of unknown parameters in Eqs.(1)--(3). Using the linear programming method \cite{MDI1,MDI3}, Alice and Bob can easily estimate the lower bound of $Y_{11}^{ZZ}$ according to Eq.(1). To get the lower bound of $g_{11}$ (${g_{11}} = C_{11}^{QS} + C_{11}^{RS} + C_{11}^{RT} - C_{11}^{QT}$), they need to estimate the upper bound of $C_{11}^{QT}$ and the lower bound of $C_{11}^{QS}$, $C_{11}^{RS}$, and $C_{11}^{RT}$ according to Eqs.(2)--(3). For example, as shown in \cite{MDI1,MDI3}, to get the upper bound of $C_{11}^{QT}$, they divide the upper bound of ${\sum\nolimits_{i,j = 0}^1 {Y_{11}^{ij,QT}C_{11}^{QT}} }$ with the lower bound of ${\sum\nolimits_{i,j = 0}^1 {Y_{11}^{ij,QT}} }$. Similarly, to get the lower bound of $C_{11}^{QS}$ ($C_{11}^{RS}$ or $C_{11}^{RT}$), they divide the lower bound of ${\sum\nolimits_{i,j = 0}^1 {Y_{11}^{ij,QS}C_{11}^{QS}} }$ (${\sum\nolimits_{i,j = 0}^1 {Y_{11}^{ij,RS}C_{11}^{RS}} }$ or ${\sum\nolimits_{i,j = 0}^1 {Y_{11}^{ij,RT}C_{11}^{RT}} }$) with the upper bound of ${\sum\nolimits_{i,j = 0}^1 {Y_{11}^{ij,QS}} }$ (${\sum\nolimits_{i,j = 0}^1 {Y_{11}^{ij,RS}} }$ or ${\sum\nolimits_{i,j = 0}^1 {Y_{11}^{ij,RT}} }$). With the estimated parameters, the final key rate is given by \cite{Lo,MDI1,Hongwei}
\begin{equation}
\begin{array}{l}
 R \ge P_{11}^{{\mu _s}{\nu _s}}Y_{11}^{ZZ}(1 - {\log _2}(1 + \sqrt {2 - \frac{{g_{11}^2}}{4}} )) \\
 \ \ \ \ \ \  - Q_{{\mu _s}{\nu _s}}^{ZZ}fh(E_{{\mu _s}{\nu _s}}^{ZZ}) \\
 \end{array},
\end{equation}
where $\mu_s$ ($\nu_s$) denotes the intensity of Alice's (Bob's) signal state, $f$ is the key reconciliation efficiency, ${E_{{\mu _s}{\nu _s}}^{ZZ}}$ denotes the error rate with intensities $\mu_s$ and $\nu_s$ in $ZZ$, and $h(x) =  - x{\log _2}(x) - (1 - x){\log _2}(1 - x)$.

{\it Simulation  - }Assume that the dark count rate of the single photon detector is $6 \times {10^{ - 6}}$, the detection efficiency of the single photon detector is 14.5\%, the loss coefficient of the quantum channel is 0.2 dB/km, and the key reconciliation efficiency is 1.16. For simplicity and without loss of generality, we only consider the measurement of the projection into the Bell state $\left| {{\psi ^ - }} \right\rangle  = \frac{1}{{\sqrt 2 }}(\left| {HV} \right\rangle  - \left| {VH} \right\rangle )$ for the CHSH-MDI-QKD protocol and the MDI-QKD protocol.

Using these parameters, we investigate the decoy-state CHSH-MDI-QKD protocol and the decoy-state MDI-QKD protocol. For the decoy-state CHSH-MDI-QKD protocol with three, four, and five intensities, the intensities of the decoy states are reasonably adopted as $\{ {\mu _0} = {\nu _0} = 0,{\mu _1} = {\nu _1} = 0.01\} $, $\{ {\mu _0} = {\nu _0} = 0,{\mu _1} = {\nu _1} = 0.01,{\mu _2} = {\nu _2} = 0.02\} $, and $\{ {\mu _0} = {\nu _0} = 0,{\mu _1} = {\nu _1} = 0.01,{\mu _2} = {\nu _2} = 0.02,{\mu _3} = {\nu _3} = 0.03\} $. For the decoy-state MDI-QKD protocol with three and four intensities, the intensities of the decoy states are the same as the decoy-state CHSH-MDI-QKD with three and four intensities. The signal state (${\mu _s}={\nu _s}$) is optimized by searching with a step of 0.01 for all cases.
\begin{figure}[htbp]
\centerline{\includegraphics[width=\columnwidth]{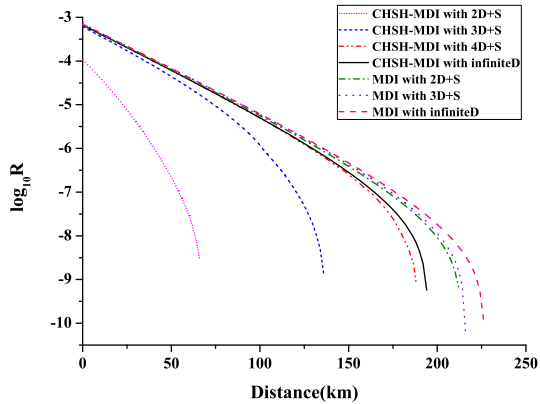}}
\caption{(Color online) Results of the decoy-state CHSH-MDI-QKD protocol and the decoy-state MDI-QKD protocol with different intensities in the asymptotic case. Note that D denotes the decoy states, and S denotes the signal state.}
\end{figure}

\begin{figure}[htbp]
\centerline{\includegraphics[width=\columnwidth]{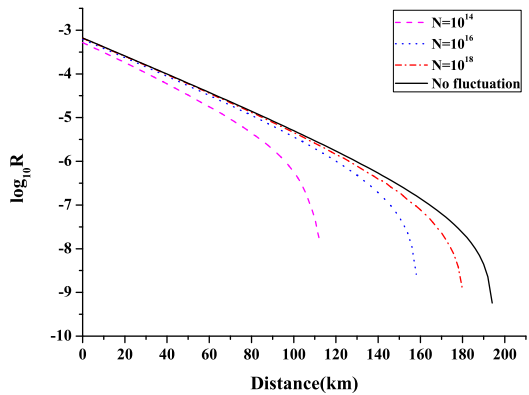}}
\caption{(Color online) Results of the decoy-state CHSH-MDI-QKD protocol with five intensities with statistical fluctuation.}
\end{figure}
The simulation results of the two protocols in the asymptotic case are presented in Fig.1, where the four lines on the left are simulation results of the decoy-state CHSH-MDI-QKD protocol, and the three lines on the right are the simulation results of the decoy-state MDI-QKD protocol. The results show that the MDI-QKD protocol with three and four intensities can achieve good approximations of the infinite intensities case, while the CHSH-MDI-QKD protocol with three and four intensities cannot, which can be explained by the fact that there are more parameters to be estimated in CHSH-MDI-QKD than in MDI-QKD using the linear programming method. And CHSH-MDI-QKD with five intensities can achieve a good approximation of the infinite intensities case.

We also consider the finite-size effect on the decoy-state CHSH-MDI-QKD protocol with five intensities, where five standard deviations of fluctuation are used\cite{decoy2}. Assume that the pulse number of decoy states and the signal states of Alice and Bob are the same, denoted by $N$. The results are shown in Fig. 2. As illustrated in Fig. 2, the secure key rate varies greatly with different pulse numbers. The secure distance with $N = {10^{14}}$ is more than 110 km.

{\it Conclusion  - }In conclusion, the CHSH-MDI-QKD protocol weakens the assumption of state preparation in MDI-QKD, and we have shown the feasibility of the decoy-state CHSH-MDI-QKD protocol with three, four, and five intensities. Especially, the decoy-state CHSH-MDI-QKD protocol with five intensities can achieve a good approximation of infinite intensities, which is very promising and can be adopted to practical QKD systems with current technology.

{\it Acknowledgments  - }This work was supported by the National Basic Research Program of China (2011CBA00200 and 2011CB921200), the National Natural Science Foundation of China (61101137, 61201239, 61205118, and 11304397), and China Postdoctoral Science Foundation (2013M540514). $^a$lihw@mail.ustc.edu.cn, $^b$yinzheqi@mail.ustc.edu.cn.

\end{document}